\documentclass[sigconf,9pt, preprint]{acmart}
\usepackage[utf8]{inputenc}

\usepackage{graphicx}
\usepackage{enumitem}
\usepackage{cleveref}
\usepackage{soul}
\usepackage{blkarray}

\title{Identifying Grey-box Thermal Models with Bayesian Neural Networks}

\author{Md Monir Hossain}
\affiliation{%
  \institution{University of Alberta}
  \city{Edmonton} 
  \state{Alberta}}
\email{mdmonir@ualberta.ca}

\author{Tianyu Zhang}
\affiliation{%
  \institution{University of Alberta}
  \city{Edmonton} 
  \state{Alberta}}
\email{tzhang6@ualberta.ca}

\author{Omid Ardakanian}
\affiliation{%
 \institution{University of Alberta}
 \city{Edmonton}
 \country{Canada}}
\email{ardakanian@ualberta.ca}

\renewcommand{\shortauthors}{Hossain et al.}

\setcopyright{none}

\acmDOI{}

\acmISBN{}

\acmConference[arXiv preprint]{arXiv preprint}
\acmYear{2020}
\copyrightyear{2020}

\makeatletter
\def\@copyrightspace{\relax}
\makeatother

\begin{document}

\begin{abstract}
Smart thermostats are one of the most prevalent home automation products.
They learn occupant preferences and schedules, 
and utilize an accurate thermal model to reduce 
the energy use of heating and cooling equipment 
while maintaining the temperature for maximum comfort.
Despite the importance of having an accurate thermal model 
for the operation of smart thermostats,
fast and reliable identification of this model is still an open problem.
In this paper, we explore various techniques for establishing a suitable thermal model
using time series data generated by smart thermostats.
We show that Bayesian neural networks can be used 
to estimate parameters of a grey-box thermal model
if sufficient training data is available,
and this model outperforms several black-box models in terms of the temperature prediction accuracy.
Leveraging real data from 8,884 homes equipped with smart thermostats,
we discuss how the prior knowledge about the model parameters can be utilized
to quickly build an accurate thermal model for another home 
with similar floor area and age in the same climate zone.
Moreover, we investigate how to adapt the model originally built for the same home in another season
using a small amount of data collected in this season.
Our results confirm that maintaining only a small number of pre-trained thermal models
will suffice to quickly build accurate thermal models for many other homes, 
and that 1~day smart thermostat data 
could significantly improve the accuracy of transferred models in another season.
\end{abstract}




\maketitle

\section{introduction}

The demand for smart and programmable thermostats has been on the rise in the past decade
thanks to improving living standards in developing countries,
and increasing retail electricity prices around the world.
The global smart thermostat market was pegged at \$1.36 billion in 2018
and is anticipated to reach \$8.78 billion by 2026~\cite{smartterm}.
Smart thermostat devices, such as ecobee, Nest, and Resideo,
take into account their measurement of environmental variables and building occupancy 
along with weather forecasts to optimally control heating and cooling equipment 
while maintaining the room temperature within desired limits.
This optimal control could result in significant energy and cost savings in buildings, 
which are responsible for around 40\% of the global energy use.
For example, ecobee claims that eco+ customers can save up to 23\% 
on their heating and cooling costs~\cite{ecobeesavings}
which is on par with cost savings reported by Nest, 
i.e., 10\% on heating and 15\% on cooling~\cite{nestsavings}.

To maintain a healthy and comfortable living and working environment,
smart thermostats rely on a black-box or grey-box heat transfer model. 
This model relates changes in the room temperature 
to a set of exogenous and control variables, e.g., 
ambient temperature, and the amount of heat injected into or extracted 
from the space by the heating or cooling system~\cite{Kleiminger2014,Salakij2016}.
The lumped parameter models --- aka Resistance-Capacitance (RC) network models 
for the analogy between heat flux and power flow~\cite{Harish2016} ---
are one of the most popular grey-box thermal models due to their simplicity.
They divide the envelope and interior of a building 
into a number of temperature-uniform \emph{lumps}
and describe temperature dynamics inside the building 
using linear differential equations derived from the heat transfer theory.


While the accuracy of a thermal model may increase with its order, 
training a high-order model is computationally expensive.
Reduced-order thermal models can strike a balance between accuracy and model complexity~\cite{Gouda2002}.
However, identifying even a low-order thermal model (e.g., 2R2C or 3R3C) requires several days of data.
Should this data be available, the parameters of the RC model can be estimated
by solving a constrained optimization problem~\cite{Harish2016,Gouda2002},
using a Kalman filter~\cite{maasoumy2014handling}
or through supervised learning~\cite{Pathak2019}.
While sufficient training data is being gathered by a smart thermostat to identify the building's thermal model,
the thermostat may take suboptimal control decisions 
which violate the thermal comfort requirements and increase the energy use.
Thus, reducing the amount of training data and the time necessary for 
building an accurate thermal model is an important research problem which we study in this paper.

Leveraging data from smart thermostats installed in 8,884 homes in Canada,
we explore various techniques for establishing a well-suited grey-box thermal model for each home.
We show that a first-order model, i.e., 1R1C, does not yield an acceptable level of accuracy 
in predicting indoor temperature for many homes in this dataset.
This motivates us to build a 2R2C model for a small number of homes (chosen by clustering)
using several days of training data collected during one season.
We utilize a Bayesian Neural Network (BNN) to estimate parameters of the RC model,
and investigate how to transfer the prior knowledge about the model parameters
to another season and across homes that belong to the same cluster.
The clusters are formed using metadata (floor area and year built),
which is readily available for the homes in our dataset.
This allows us to select the most appropriate model from a library of pre-trained models
for a given home knowing only its age and floor area, 
and eliminates the need to collect several days of training data from each home.
Thus, from the first day, the smart thermostat can use a fairly accurate thermal model 
to optimize comfort and energy use.

The main intuition for this work is derived from the observation that
\emph{despite all differences in the layout and design of homes, 
there is a limited number of combinations of RC parameters}.
This is especially true, if we restrict our focus to one country 
with a narrow range of climates and specific building codes.
Thus, the knowledge acquired in one home will be quite useful in another similar home.\\
Our contribution is threefold:
\begin{itemize}[noitemsep,topsep=0pt]
    \item We propose a methodology based on Bayesian neural networks
    for identifying the RC model  
    of a home equipped with a smart thermostat.
    We compare the predictive power of different RC models
    and show that a 2R2C model yields lower accuracy than other low-order models 
    for most homes in our dataset.
    
    \item We show that a grey-box model is more accurate than several black-box models, 
    including time series and neural network models, in predicting the room temperature.
    
    
    
    \item We assign homes in our dataset to a small number of clusters based on 
    their floor area and age, and show that this clustering allows 
    for transferring a pre-trained representative model of that cluster 
    to this home with and without adaptation. 
    We also discuss how a model trained for one season can be transferred to another season.
    
\end{itemize}

The dataset we use in this study contains time series data and metadata 
obtained from a large number of \emph{ecobee} smart thermostats.
We describe this dataset in Section~\ref{sec:dataset}.

\section{Related Work}

The optimal control of the building Heating, Ventilation, and Air Conditioning (HVAC) system
is of great importance as it accounts for a large fraction of the building energy use, 
and is responsible for maintaining the temperature inside the building within a comfortable range.
To optimally control this system, most related work adopts receding horizon control 
which relies on a model that explains how the room temperature 
changes as a result of implementing a certain control policy.
This has given rise to a large number of studies aiming to 
solve a \emph{system identification} problem to infer this thermal model.

A variety of data-driven techniques have been used 
in the literature to establish the thermal model. 
This includes Artificial Neural Network (ANN)~\cite{pollard1998occupant}, 
ANN with Levenberg-Marquardt (LM)~\cite{moon2013development, kim2014performance, moon2016algorithm}, 
ANN with Backpropagation (BP)~\cite{moon2009application, buratti2014building}, 
Multilayer Perceptron (MLP) with LM~\cite{mba2016application, huang2015neural}, 
Radial Basis Function (RBF)~\cite{ruano2006prediction, ferreira2002choice}, 
Autoregressive model with exogenous variables (ARX)~\cite{mustafaraj2011prediction, lu2009prediction}, and
Autoregressive Moving Average model with exogenous inputs (ARMAX)~\cite{mustafaraj2010thermal, patil2008modelling}.
These black-box models essentially map a number of features to the room temperature,
and their accuracy highly depends on the selected features~\cite{enescu2017review}.


Another type of thermal models is the RC model 
which is commonly used for heat transfer analysis in buildings.
This grey-box model turns building spaces and multi-layered walls 
into a number of latent thermal resistances and capacitances.
Despite its simplicity, it achieves a high accuracy in predicting the indoor temperature.
Zhou et al.~\cite{zhou2017quantitative} compare a low dimensional RC model with a physics-based model, 
and conclude that the RC model can substitute the physics-based model with a negligible loss of accuracy. 

The RC model can be arbitrarily complex. 
Several attempts have been made to date to represent the building interior and its envelope
using RC network models of different orders.
For example, 4R1C~\cite{maasoumy2014handling, haldi2010impact}, 
3R2C~\cite{ogunsola2014development, zhu2011energy}, and 2R1C~\cite{Gouda2002} 
networks have been used to model the building envelope, 
while 1C~\cite{maasoumy2014handling}, 1R1C~\cite{haldi2010impact, zhu2011energy}, and 
2R2C~\cite{ogunsola2014development} networks have been used to represent the building spaces.
The majority of studies that build an RC model 
assume the knowledge of the building insulation, 
thermal mass, floor area, layout, and construction material~\cite{fayazbakhsh2015resistance, braun2002inverse, seem1989transfer, deconinck2016comparison}. 
Leveraging this knowledge can indeed greatly reduce the complexity of model training. 
However, this contextual and physical information is typically hard to obtain without an intrusive energy audit, 
especially for residential buildings that we study in this work. 
To overcome this barrier, one approach is to estimate the parameters of the RC model 
from time series data generated by smart thermostats. 
Specifically, the model can be built by solving a constrained optimization problem~\cite{Harish2016, Gouda2002}, 
employing an unscented Kalman filter~\cite{maasoumy2014handling}, 
and utilizing the genetic algorithm~\cite{ogunsola2014development}. 
All these methods require a significant amount of data to build an accurate model. 
On the contrary, this paper focuses on building a suitable RC model 
for a given residential building utilizing its metadata and a small amount of time series data.

Our work is similar to~\cite{baasch2019comparing}, 
which focuses on estimating the amount of heat loss through the building envelope using a 1R1C model. 
The authors evaluate the balance point, decay curve, and energy balance models
to cluster buildings and estimate the heat loss. 
Compared to that work, we form clusters based on metadata,
determine the level of complexity of the RC model for each building,
and use transfer learning to reduce the amount of data needed for system identification. 
Pathak et al.~\cite{Pathak2019} use Bayesian learning to estimates the RC parameters 
for one season and then utilize the learned parameters in another season. 
Our work is similar to that work
but we explore the possibility of adapting and reusing an RC model in buildings
that have similar characteristics to the building for which this model was initially built.

Transfer learning allows for taking advantage of the knowledge obtained in one model in another model. 
For example, Zhang et al.~\cite{zhang2019domain} employ transfer learning 
to transfer an occupancy estimation model across different buildings. 
Although transfer learning can provide initial estimates for the model parameters,
there are other challenges such as over-fitting and unobserved patterns in the training data of the source domain.
Bayesian neural networks include uncertainty in the weights and biases 
to resolve this issue~\cite{blundell2015weight}. 
The uncertainty in neural networks avoids over-fitting and 
makes it possible to identify unseen patterns and give reasonable predictions~\cite{lauret2008bayesian}. 
A recent study utilized BNN together with transfer learning to build thermal models~\cite{hossain2019evaluating}. 
Compared to that work, we discuss how to choose the order/complexity of the RC model, 
compare this RC model with several black-box models proposed in the literature,
and incorporate metadata in the model training process to reduce the need for temperature time series data.
We also investigate how to transfer a model trained in one season to another season within the same home.
\section{Dataset}\label{sec:dataset}
To build and evaluate grey-box and black-box thermal models, 
we use the smart thermostat dataset released by \emph{ecobee} -- 
one of the key players in the smart thermostat market -- 
as part of a program called `donate your data'. 
This program allows customers to voluntarily share 
their anonymized smart thermostat data with researchers to foster research and development. 
The dataset contains time series data generated by smart thermostats 
along with metadata about homes where the thermostats are installed. 
The metadata contains information about the homes and their occupants, 
e.g., location, year built, floor area, typical occupancy level, 
number of cooling and heating stages, etc.

\begin{table}[t!]
\caption{A subset of time series data included in the dataset.}\label{tab:dataset}
\begin{tabular}{@{}ccc|ccc@{}}
\toprule
Feature              & Data Type & Unit        & Feature   & Data Type & Unit     \\ \midrule
$\text{HVAC}_{mode}$ & Text      & N/A         & $T_{out}$ & Decimal   & $^{\circ}F$ \\
$T_{in}$             & Decimal   & $^{\circ}F$ & $Motion$  & Binary    & N/A       \\
$T_{setcool}$        & Decimal   & $^{\circ}F$ & $H_{in}$  & Decimal (pct.)   & N/A \\
$T_{setheat}$        & Decimal   & $^{\circ}F$ &&&\\\bottomrule
\end{tabular}
\end{table}

There is a total of 104,693 homes in this dataset
located in 51 countries with different climates and building codes.
In this paper, we just consider homes that are located in Canada 
and are therefore in the same climate zone (Zone 7).
This yields 12,960 homes which is a reasonable size sample.
The data is available between 2015 to 2019. 
The number of participating homes has increased over time 
as more homes equipped with ecobee thermostats opted in to participate in this program.
Hence, the lengths of time series data are different for different homes. 
Moreover, the number of homes varies based on the season. 

We initially restrict our study to winter due to the additional complexity 
that mixing data from different seasons presents~\cite{Pathak2019}. 
For the winter season, we consider homes in our dataset that have 
at least 3 months worth of data between November 2018 to February 2019. 
This gives us 8,884 homes to run the experiment on.
We then attempt to transfer models from the winter season to the summer season 
considering homes that have reported at least 3 months worth of data between May and August 2019. 
This gives us 8,834 homes to run the experiment on.

We identify and separate the homes located in Canadian using the metadata file 
that contains information regarding the country, province, and city a specific home is located in. 
We also use a subset of time series data which exist in this dataset. 
In particular, for each home we use indoor and outdoor temperatures, humidity, 
motion indicator, operation mode of heating and cooling equipment, heating setpoint, and cooling setpoint. 
All these features are recorded at 5-minute intervals.
We note that this dataset does not contain solar radiation, 
energy use of plug loads, or more accurate or fine-grained occupancy information. 
Thus, we cannot incorporate them in our model. 

Table~\ref{tab:dataset} shows the features used in this study.
Here, HVAC\textsubscript{Mode} represents the state of the HVAC system, 
which can be one of the following textual values: `off', `heat', `cool', and `auto'. 
$T_{in}$ is the average indoor temperature. 
$T_{setcool}$ and $T_{setheat}$ are indoor cooling and heating setpoints, respectively. 
$T_{out}$ is the outdoor temperature that is collected from the nearest weather station. 
$H_{in}$ is the relative humidity inside the building and motion
is a binary variable which is 1 when motion is detected by the sensor and is 0 otherwise.
To impute missing data (except for the motion data) we use a simple linear interpolation technique. 
For motion data, we set all missing values to zero.



\section{Grey-box Thermal Models}\label{sec:background}
In this section we explain a grey-box modelling approach which builds a thermal model
for a building given time series data gathered by the thermostat
installed in that home.

\begin{figure}[t!]
\centerline{\includegraphics[width=.7\columnwidth]{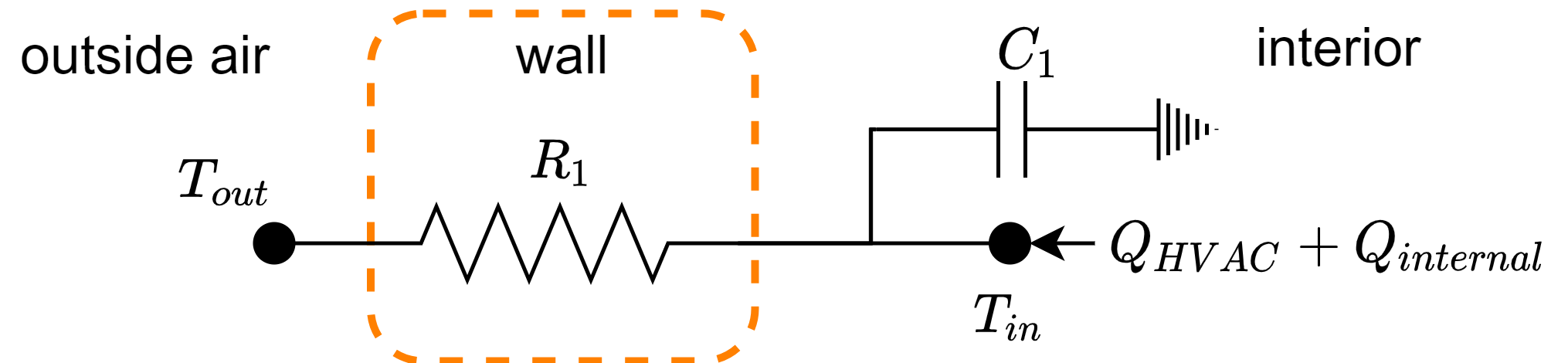}}
\caption{A first-order RC model representing the building envelope with a single thermal resistor (1R) 
and the building interior with a thermal capacitor (1C).}
\label{fig:1r1c}
\end{figure}

\begin{figure}[t!]
\centerline{\includegraphics[width=\columnwidth]{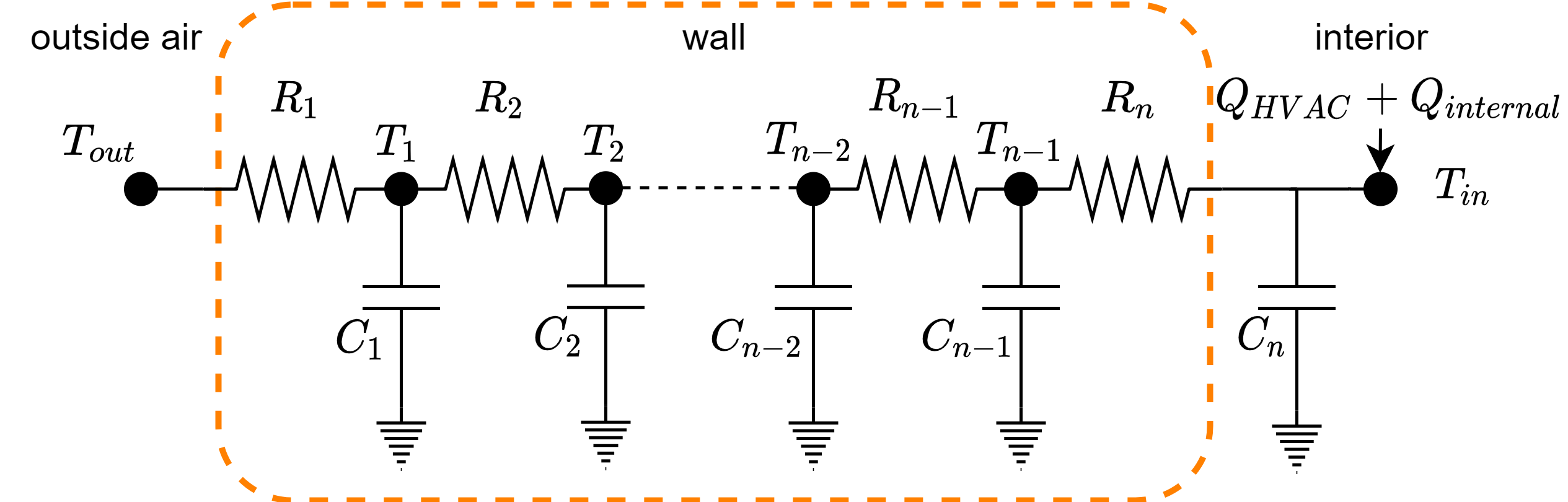}}
\caption{An RC model representing the building envelope with an $n$R$(n-1)$C network
and the interior of the building using a thermal capacitor (1C).}
\label{fig:3r3c}
\end{figure}

The RC model uses the analogy between thermal and electrical conduction 
to model heat flux, $Q$, in a building. 
It models the heat transfer across the building envelope and inside the building 
using multiple constant thermal resistors and capacitors 
connecting spatially temperature-uniform lumps.
Specifically, the building exterior and interior can be modelled 
by a number of constant resistances and capacitances.
The thermal resistor reduces the current heat flux between its two terminals. 
The amount of heat that can pass through the resistor can be computed from $Q=\Delta T/R$,
where $\Delta T$ is the temperature difference between the two terminals and $R$ is the thermal resistance. 
The thermal capacitor stores heat according to $C\frac{\partial T}{\partial t}=Q$, 
where $C$ represents the constant capacitance.

To build an RC model, the first step is to decide on its order (determining the model complexity). 
In general, the higher the complexity is, the more accurate the model becomes
as it does not lump together elements that do not have the same temperature.
However, increasing the order of the model is a mixed blessing.
The higher the order of the model is the more difficult 
it becomes to estimate the RC parameters in terms of computational cost and the amount of training data needed.
More information would also be required about 
different spaces within the building (e.g., the blueprint) to train a high-order RC model.
Prior work suggests that a 2nd-order RC model gives a negligible loss of accuracy compared 
to a 20th-order RC model~\cite{Gouda2002}. 
Hence, we focus on building an accurate low-order RC model for residential buildings.

\noindent \textbf{$n$R$n$C Models:}
We model the building envelope using an $n$R$(n-1)$C network
since new built homes typically have multi-layer walls and insulated glazing.
When $n=1$ the 1R model lumps the building envelope into one thermal resistor, denote by $R_1$.
The $n$R$(n-1)$C model for $n>1$ divides the building envelope into $2n-1$ components, 
including $n$ resistors, denoted by $R_1, \cdots, R_n$, and 
$n-1$ capacitors, denoted by $C_1, \cdots, C_{n-1}$.
This model can represent the $n$-layer structure of the wall~\cite{Harish2016}.
The $2n-1$ parameters depend on the thickness, materials used, and insulation of the wall.

We use a single capacitor to represent the building interior as one thermal zone which absorbs and retains heat.
The constant capacitance is denoted by $C_1$ in the 1R1C model and $C_n$ in the $n$R$n$C model.
This thermal capacitance determines how much inertia the building provides against temperature fluctuations, 
and depends on the building's floor area and ceiling height.
Hence, the larger a building is, the higher its thermal mass (or capacitance) would be.

Since the heating (or cooling) system can inject heat into (or extract heat from) the space,
we connect it to the node that represents the building space.
We denote by $Q_{HVAC}$ the heat flux caused by HVAC
and denote by $Q_{internal}$ the internal heat gain due to occupant presence and activities,
and other latent variables (e.g., appliance usage).
Figure~\ref{fig:1r1c} and Figure~\ref{fig:3r3c} 
illustrate the 1R1C and $n$R$n$C models, respectively.

The temperature dynamics in the 1R1C model can be described by the following equation:
\begin{equation}
\begin{aligned}
\partial T_{in} = \frac{1}{C_1R_1}(T_{out}-T_{in})\partial t + \frac{Q_{HVAC}+Q_{internal}}{C_1}\partial t
\end{aligned}
\label{eq:1r1c}
\end{equation}
where $T_{in}$ represents the temperature inside the building, $T_{out}$ represents the outside temperature,
and $Q_{HVAC} = k_{heat}Q_{heat} - k_{cool}Q_{cool}$ where $k_{heat}$ and $k_{cool}$ are two binary control variables determined by the operation mode of the HVAC system, i.e., $\text{HVAC}_{\text{mode}}$:
\begin{align*}
\begin{split}
k_{heat}&=
\begin{cases}
1, & \text{if}~T_{in} < T_{setheat}~\&~\text{HVAC}_\text{mode}\in\{auto,heat\}\\
0, & \text{otherwise}  
\end{cases}\\
k_{cool}&=
\begin{cases}
1, & \text{if}~T_{in} > T_{setcool}~\&~\text{HVAC}_\text{mode}\in\{auto,cool\}\\
0, & \text{otherwise}  
\end{cases}
\end{split}
\end{align*}
This simplified model incorporates the heat introduced or extracted by the HVAC system and 
the internal heat gain due to occupancy, but neglects solar radiation, and different wall insulation layers.

Similarly, the $n$R$n$C model can be expressed using a system of linear differential equations 
describing heat flow inside the building and across its envelope:
\begin{align}
\nonumber \partial T_1 &= \frac{1}{C_1R_1}(T_{out}-T_1)\partial t + \frac{1}{C_1R_2}(T_2-T_1)\partial t \\
\nonumber \partial T_i &= \frac{1}{C_iR_i}(T_{i-1}-T_i)\partial t + \frac{1}{C_iR_{i+1}}(T_{i+1}-T_i)\partial t\quad {\scriptstyle \forall i\in\{2,\dots,n-1\} } \\
\label{eq:3r3c} \partial T_{in} &= \frac{1}{C_nR_n}(T_{n-1}-T_{in})\partial t + \frac{Q_{HVAC}+Q_{internal}}{C_n}\partial t 
\end{align}
where $T_i$'s are the envelope temperatures as depicted in Figure~\ref{fig:3r3c}. 
This model incorporates the heat introduced or extracted by the HVAC system and internal heat gain due to occupancy,
but neglects the heat flux from solar radiation
as solar radiation and the window area are not captured in our dataset.

In both models, we assume that the temperature sensor (which is embedded in the thermostat) 
is located in a place that can measure the indoor temperature with negligible error. 
Moreover, we use constant values $Q_{heat}$ and $Q_{cool}$ to represent the 
heat injected or extracted by the HVAC system,
assuming that it injects and extracts heat at constant rates in the heating and cooling stages.
This assumption does not necessarily hold in practice; nevertheless, 
our experiments show that it does not introduce a significant error. 

We also assume that $Q_{internal} = f(\text{motion}, \text{humidity})$
where $f$ is any function of motion and relative humidity, 
the quantities that are measured directly by smart thermostats.
Unfortunately, the binary motion data is not typically a good proxy for home occupancy
as a home might be occupied during a period with no recorded motion events.
Our experimental results confirm that incorporating the measured motion and relative humidity
in the grey-box and black-box models described in the next section 
has an imperceptible impact on the predictive power of these models\footnote{The difference between 
the average RMSE with and without these features is 0.0066 
for 100 homes that are randomly selected from the dataset.}.
Thus, we ignore the internal heat gain due to occupant presence and activities, i.e., $Q_{internal}$,
in the rest of the paper. This simplifies the indoor heat gain or loss to $Q_{HVAC}$.
We describe how to select the order of the RC model and how to estimate model parameters 
using a Bayesian neural network in Section~\ref{sec:methodology}.

\section{Methodology}\label{sec:methodology}
In this section we discuss how to estimate parameters of an RC model 
and how to train various black-box thermal models for a given home
assuming that a sufficient amount of smart thermostat time series data is available.



\subsection{Parameter Estimation for RC Models\label{sec:rc-model}}
To estimate the model parameters 
we first transform the linear differential equations to linear difference equations.
This allows us to use different techniques for model parameter estimation
considering several consecutive measurements from the smart thermostat.



\subsubsection{1R1C model}
We transform the linear differential equation~(\ref{eq:1r1c}) to a linear difference equation
by replacing the temperature difference between two consecutive time slots with $\Delta T_{in}$ 
and the temperature difference between the outdoor and indoor environment with $\Delta T_{diff}$:
\begin{align*}
\Delta T_{in} &= \frac{1}{C_1R_1}\Delta T_{diff} + k_{heat}\cdot\frac{Q_{heat}}{C_1} - k_{cool}\cdot\frac{Q_{cool}}{C_1}
\end{align*}

Given $m$ successive values of $\Delta T_{in}$ and $\Delta T_{diff}$, and 
leveraging the fact that $R_1C_1, \frac{Q_{heat}}{C_1}, \frac{Q_{cool}}{C_1}$ cannot be negative, 
we solve the following non-negative least squares problem to determine 
the model parameters which are collected in vector $x$:
\begin{align*}
&\min_{x} ||\mathbf{A}x-y||_2 \\
&~~\text{s.t.}\qquad x\succcurlyeq 0
\end{align*}
where
\begin{equation*}
    \mathbf{A}
    =
    \begin{bmatrix}
        \Delta T^{[1]}_{diff} & k^{[1]}_{heat} & -k^{[1]}_{cool}\\[6pt]
        \Delta T^{[2]}_{diff} & k^{[2]}_{heat} & -k^{[2]}_{cool}\\[6pt]
        &\cdots&\\
        \Delta T^{[m]}_{diff} & k^{[m]}_{heat} & -k^{[m]}_{cool}
    \end{bmatrix}\hspace{1em}
    x
    =
    \begin{bmatrix}
        \frac{1}{C_1R_1} \\[10pt]
        \frac{Q_{heat}}{C_1} \\[10pt]
        \frac{Q_{cool}}{C_1}
    \end{bmatrix}\hspace{1em}
    y
    =
    \begin{bmatrix}
        \Delta T^{[1]}_{in} \\[10pt]
        \Delta T^{[2]}_{in} \\[10pt]
        \vdots\\
        \Delta T^{[m]}_{in}
    \end{bmatrix}
\end{equation*}
We use the active-set method to solve this convex optimization problem.

\subsubsection{$n$R$n$C model}
To estimate the parameters of an $n$R$n$C model,
we first find the time domain solution of the system of $n+1$ linear differential equations 
expressed in~(\ref{eq:3r3c}).
The state-space representation of this system is:
\begin{align}
\label{eq:state} \frac{\partial x}{\partial t} &= \mathbf{A}x + \mathbf{B}u,\\
y &= \mathbf{C}x + \mathbf{D}u, \label{yeq}
\end{align}
where
\vspace{-10pt}
\begin{equation*}
\mathbf{A}
=
\begin{blockarray}{cccccc}\\
    \begin{block}{[ccccc]c}
        -\left(\frac{1}{C_1R_1}+\frac{1}{C_1R_2}\right) & \frac{1}{C_1R_2} & 0 & \cdots & 0 & \text{row 1}\\[3pt]
        \BAmulticolumn{5}{c}{\xleftarrow{\hspace*{1cm}}\hspace{25pt}v_{2\phantom{-1}}\hspace{25pt}\xrightarrow{\hspace*{1cm}}} & \text{row 2}\\[-5pt]
        \BAmulticolumn{5}{c}{\vdots} & \\[-5pt]
        \BAmulticolumn{5}{c}{\xleftarrow{\hspace*{1cm}}\hspace{25pt}v_{n-1}\hspace{25pt}\xrightarrow{\hspace*{1cm}}} & \text{row }n-1\\[3pt]
        0 & \cdots & 0 & \frac{1}{C_nR_n} & -\frac{1}{C_nR_n} & \text{row }n\\
    \end{block}
\end{blockarray}
\end{equation*}
\vspace{-10pt}
\begin{equation*}
    v_{i} = \Bigg[\underbrace{0, \cdots, 0}_{i-2\text{ zeros}}
    , \frac{1}{C_iR_i}, -\left(\frac{1}{C_iR_i}+\frac{1}{C_iR_{i+1}}\right), \frac{1}{C_iR_{i+1}}, \underbrace{0, \cdots, 0}_{n-i-1\text{ zeros}}\Bigg]
\end{equation*}

\begin{equation*}
\mathbf{B}
=
\begin{bmatrix}
\frac{1}{C_1R_1} & 0 & 0\\[6pt]
0 & 0 & 0\\
\vdots & \vdots & \vdots\\
0 & 0 & 0\\[6pt]
0 & \frac{Q_{heat}}{C_n} & -\frac{Q_{cool}}{C_n}
\end{bmatrix}\hspace{3em}
\mathbf{C}
=
\begin{bmatrix}
0 \\
\vdots\\[6pt]
0 \\[10pt]
1
\end{bmatrix}^\top\hspace{3em}
\mathbf{D}
=
\begin{bmatrix}
0 \\[10pt]
\vdots \\[10pt]
0
\end{bmatrix}^\top
\end{equation*}
Here the state vector $x = \begin{bmatrix} T_1 & T_2 & \cdots & T_{n-1} & T_{in} \end{bmatrix}^\top$ 
collects envelope and room temperatures,
the output $y$ denotes the indoor temperature ($T_{in}$),
the input vector $u = \begin{bmatrix} T_{out} & k_{heat} & k_{cool} \end{bmatrix}^\top$ 
collects the outdoor temperature and the HVAC system control inputs,
and $\mathbf{A}, \mathbf{B}, \mathbf{C}$ and $\mathbf{D}$ are constant coefficient matrices.
As shown in~\cite{seem1989transfer} the response of this system can be written as:
\begin{equation}\label{eq:sol}
    x_t=(F\cdot \mathbf{I}-\Phi)^{-1}(F\cdot \Gamma_2+\Gamma_1-\Gamma _2)u_t,
\end{equation}
where $F$ denotes the forward shift operator (i.e., $F x_t= x_{t+\delta}$
where $\delta$ denotes the time step), $\Phi= e^{\mathbf{A}\delta}$ is the matrix exponential, and
\begin{equation*}
    \Gamma_1 = \mathbf{A}^{-1}(\Phi - \mathbf{I})\mathbf{B}, \hspace{5em}
    \Gamma_2 = \mathbf{A}^{-1}\left [ \frac{\Gamma_1}{\delta} - \mathbf{B}\right ].
\end{equation*}

Substituting $x_t$ in~(\ref{yeq}) yields the system output response:
\begin{align*}
\label{simpl}
    y_t&=\mathbf{C}(F\cdot \mathbf{I}-\Phi)^{-1}(F\cdot \Gamma_2+\Gamma_1-\Gamma _2)u_t + \mathbf{D}u_t,\\
    &=\mathbf{C}(F\cdot \mathbf{I}-\Phi)^{-1}(F\cdot \Gamma_2+\Gamma_1-\Gamma _2)u_t.
\end{align*}
The second equation above holds because $\mathbf{D} = \begin{bmatrix} 0 & \cdots & 0\end{bmatrix}$.

Next we can calculate the inverse of $(F\cdot \mathbf{I}-\Phi)$ matrix, 
which is equal to the adjugate of $(F\cdot \mathbf{I}-\Phi)$ divided by determinant of $(F\cdot \mathbf{I}-\Phi)$.
Assuming that this determinant is nonzero we have:
\begin{align*}
    (F\cdot \mathbf{I}-\Phi)^{-1} =\hspace{2pt}&\frac{\mathbf{M}_0F^{n-1}+\mathbf{M}_{1}F^{n-2}+\cdots+\mathbf{M}_{n-1}}{F^{n}+e_1F^{n-1}+e_{2}F^{n-2}+\cdots+e_n}
\end{align*}
where
\begin{equation*}
\begin{aligned}
    \mathbf{M}_0 &= \mathbf{I}\\
    \mathbf{M}_i &= \Phi\mathbf{M}_{i-1}+e_i\mathbf{I}
\end{aligned}
\hspace{5em}
\begin{aligned}
    e_i &= -Tr(\Phi\mathbf{M}_{i-1}) / i\\
    e_n &= -Tr(\Phi\mathbf{M}_{n-1}) / n
\end{aligned}
\end{equation*}
and $Tr()$ denotes the trace of a matrix.
Hence, the system output response can be written in a compact form as:
\begin{equation}\label{eq:rc-solution}
    y_t=\sum_{i=0}^{n}\mathbf{S}_i u_{t-i\delta}-\sum_{i=1}^{n}e_i{y}_{t-i\delta}
\end{equation}
where
\begin{align*}
    \mathbf{S}_0 &= \mathbf{C}\mathbf{M}_0\Gamma_2\\
    \mathbf{S}_i &= \mathbf{C}(\mathbf{M}_{i-1}(\Gamma_1-\Gamma_2)+\mathbf{M}_i\Gamma_2)\qquad i\in\{1,\cdots,n-1\} \\
    \mathbf{S}_n &= \mathbf{C}\mathbf{M}_{n-1}(\Gamma_1-\Gamma_2)
\end{align*}

\begin{figure}[t!]
\centerline{\includegraphics[width=\columnwidth]{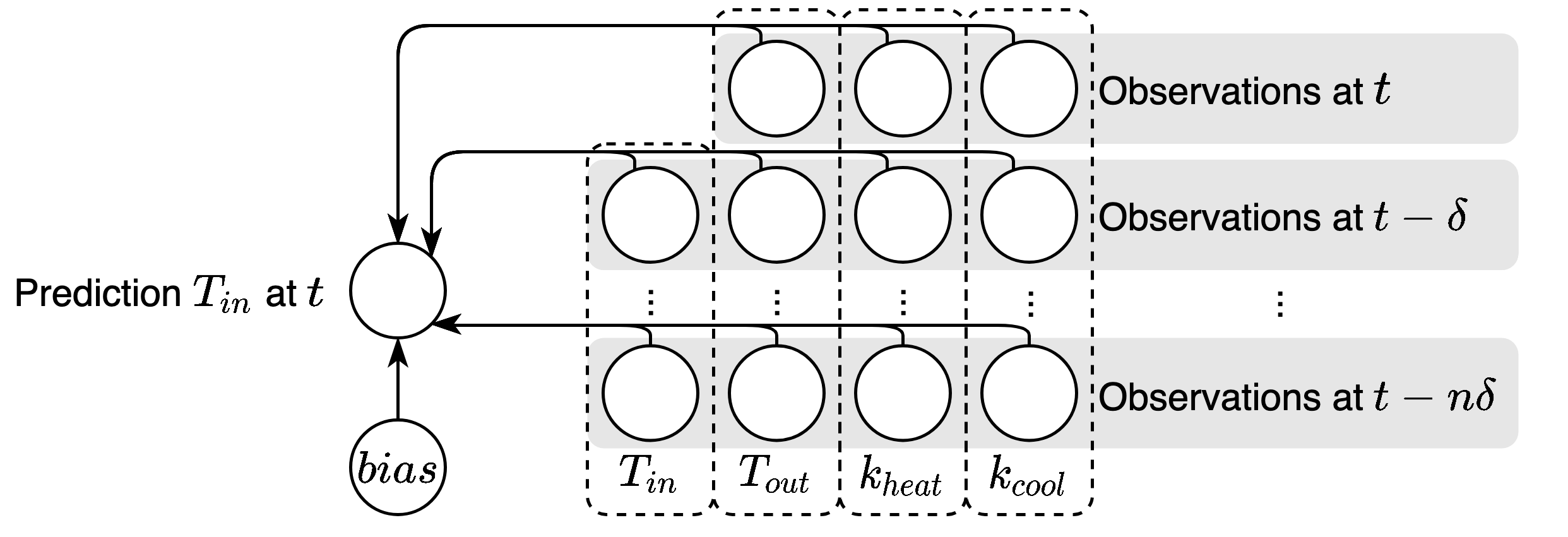}}
\caption{Structure of the BNN-RC model used to predict the indoor temperature.}
\label{fig:bnn}
\end{figure}

We get a linear function that maps 
the measured room temperature in the past $n$ time slots, and
the outside temperature and heat flux from the HVAC system in the current and past $n$ time slots 
to the room temperature in the current time slot:
$$y_t=f({u}_t, {u}_{t-1}, \cdots, {u}_{t-n},y_{t-1}, \cdots,y_{t-n})$$
Here $f$ is a linear function that can be approximated using a Bayesian neural network. 
The Bayesian neural network mimics the structure of the $n$R$n$C model; 
thus, we refer to this model as BNN-$n$R$n$C. 
Specifically, we use the terms in Equation~(\ref{eq:rc-solution}) to decide about the neurons. 
Figure~\ref{fig:bnn} illustrates the structure of our BNN-$n$R$n$C model.
We use Gaussian scale mixture distribution with $\sigma_1=1$, $\sigma_2=0.1$, and $\pi=0.2$
for the weights and biases in our BNN. 
The model learns $4n+3$ compound RC parameters for an $n$R$n$C model. 

\subsubsection{Transfer learning} 
A large amount of time series data is required to train a BNN-$n$R$n$C model
which can accurately predict the temperature inside the building.
However, sufficient training data is not always available for all homes 
and collecting this much data can be time consuming.
More importantly, the HVAC system will be controlled in a suboptimal fashion 
before a good RC model is trained and used for model predictive control.
To address this problem, we select several representative homes as source homes and 
use the posterior weights trained in those BNN-$n$R$n$C models as the prior for the new homes. 
Furthermore, we retrain the model with the small amount of data that might be available in the target home. 
We discuss the strength of transfer learning in 
Section~\ref{sec:result-transfer}.

Note that we do not transfer 1R1C models to other homes. 
This is because the 1R1C model describes the thermal resistance of the building envelope 
using a single resistor regardless of the material used, the number of layers and thickness of each layer.
Hence, the value of $R$ does not really represent any of the above factors.
Transferring this knowledge will not be helpful for training a model for a different home.

\subsection{Developing Black-Box Models}
The temperature evolution inside a home can also be modelled using a purely data-driven approach.
In this section we introduce 4 models adopted in related work.
These models can be trained using time series generated by the smart thermostat,
namely Autoregressive Integrated Moving Average with exogenous variables (ARIMAX),
Long Short Term Memory (LSTM), Deep Neural Network (DNN), and Random Forest (RF).
These models are capable of predicting the indoor temperature using the same features as the BNN-$n$R$n$C model.
We use these models as baselines to evaluate the grey-box RC model 
in the source domain and after transferring to the target domain.

\subsubsection{ARIMAX model}
The temperature time series data exhibits temporal correlation and responds 
to changes in several variables including the ambient temperature, 
and the operation mode of the HVAC system, i.e., whether it is in the heating or cooling stage.
Thus, to model the temporal correlation and capture the effects of external variables on the room temperature,
we fit an Autoregressive Integrated Moving Average model with exogenous variables (ARIMAX)~\cite{chatfield2003analysis}
to this data. 
The ARIMAX model is basically a variant of the widely used Autoregressive Moving Average (ARMA) model, 
which is more appropriate for modelling non-stationary time series
and can account for external variables which could affect the time series.
This model has three parts, namely
the autoregressive part (AR), the integrated part (I), and the moving average part (MA). 
These parts are characterized by three parameters, denoted by $p, d, q$, respectively.

The parameter $d$ of the integrated part of the model denotes the number of times we apply the differencing operator 
on the time series to ensure that the resulting time series is stationary.
Hence, we need to analyze the stationarity of the time series for estimating $d$.  
We use the Augmented Dickey-Fuller (ADF) test -- a formal statistical test for stationarity -- 
to check stationarity of the differenced time series. 
This test enables us to identify if the time series has a unit root (i.e., a stochastic trend in the time series). 



Running the ADF test on temperature data of one sample home 
reveals that the p-value is greater than 0.05.
Hence, we need to difference the time series at least once. 
We find that after one differencing the time series becomes stationary (p-value is less than 0.05).
Therefore, we set $d$ to 1.

To estimate $p$, the parameter of the autoregressive part of the ARMIAX model,
we draw the partial autocorrelation function (PACF) which measures the correlation 
between the original time series and the lagged time series 
after removing the effects of intervening past observations. 
The PACF plot of the temperature time series suggests that $p$ should be set to 1.

To estimate $q$, the parameter of the moving average part of the ARMIAX model, 
we examine the autocorrelation function (ACF) which measures 
how a time series is correlated with itself at different lags.
The ACF plot of the temperature time series suggests that $q$ should be set to 2 for the ARIMAX model.

Although using the same parameters for all homes may not result in optimal ARIMAX models, 
but it significantly reduces the complexity of building the time series model for each home.
We selected one home randomly and performed grid search 
with 125 parameter combinations to find the best ARIMAX model. 
For this specific home, the best ARIMAX model found by grid search is $ARIMAX(0, 1, 2)$ 
which has RMSE of 0.483 degree Fahrenheit, 
while using the same parameters for all ARIMAX models (i.e., $ARIMAX(1, 1, 2)$) yields RMSE of 
0.487 degree Fahrenheit for this home. 
We also tested the parameters on multiple randomly selected homes and obtained similar RMSE values.
This means that the ARIMAX models obtained using the same parameters
achieve sufficiently high accuracy in predicting the temperature inside the homes.
Hence, in this work we use the same parameters $(1, 1, 2)$ for all ARIMAX models.

\subsubsection{Long Short Term Memory (LSTM) model}
LSTM model is another time series model that is widely used to learn long-term dependencies in data. 
In particular, LSTM maintains a state to capture the history of the input and output sequence, 
enabling it to learn complex temporal dependencies. 
The LSTM network used for indoor temperature estimation is a stacked LSTM model. 
It has two hidden layers which contain 20 cells each.
This stacked model gives the best average RMSE result 
for 5 randomly sampled homes in our dataset over 20 different stacked and single-layer LSTM models.

Our LSTM model contains one output node in the output layer representing the indoor temperature. 
The input layer has 4 cells which are the previous indoor temperature, 
the current outdoor temperature, the current HVAC cooling operation state, and the current HVAC heating operation state. 
The loss is computed using the mean squared error between logits and labels. 
For each batch of the data, we minimize the loss 
using a first-order gradient-based optimization method~\cite{kingma2014adam}.
This LSTM model is fed data from $n$ consecutive time steps as a sequence, 
where $n$ depends on the order of the respective BNN-$n$R$n$C model 
(for fair comparison). 
Our experiments show that the number of time steps has a negligible impact 
on the estimation result in terms of RMSE.

\subsubsection{Deep Neural Network (DNN) model}
Deep neural networks have been successfully used in the past for forecasting temperature~\cite{romeu2013time}. 
Our DNN model consists of 400 hidden cells which are divided evenly into two hidden layers. 
It adopts Rectified Linear Unit (ReLU) as the activation function for all hidden layers to have low computational cost.
The linear activation function is used in the output layer, which contains one output node
for estimating the indoor temperature. 
This model takes the same number of input features as the benchmarking BNN-$n$R$n$C model, 
i.e., $4n+3$ features to learn the temporal correlations of the data over time. 
This architecture is determined by performing grid search over 5 different depths and 10 different numbers of units in each hidden layer. 
The DNN model utilizes the same loss function and is trained using the same optimization method as the LSTM model.

\subsubsection{Random Forest}
Random forest can determine the feature importance 
by using different subsets of available features to construct decision trees. 
Each decision tree learns multiple binary rules based on the input data. 
Random forest takes the average of estimations from all decision trees and returns it as the final result.

Random forest is computationally inexpensive, easy to implement and also powerful. 
It gives promising results for time-series prediction~\cite{kane2014comparison}. 
We use the bisection method to find the most appropriate number of trees, which is set to 100 
based on results obtained from 5 randomly selected homes. 
We use the Gini impurity measure to decide on the optimal split.
\section{Evaluation}
We build several grey-box and black-box models for the homes in our dataset and 
compare them in terms of temperature prediction accuracy in the test data set
using the root-mean-square error (RMSE). 
We answer the following questions in this section:
(a) Which RC model has the best overall performance considering all homes in our dataset? 
(b) How much training data is needed to build an accurate model? 
(c) Can we utilize a pre-trained RC model with little or no adaptation (retraining) 
to estimate the temperature inside a given home? 
(d) Can we transfer RC models across seasons? 
(e) Does a Bayesian neural network offer any advantage over a standard neural network
in the estimation of RC model parameters?
(f) How does the BNN-RC model perform compared to black-box models?

\subsection{Selecting the order of RC models}
Determining the most appropriate order of RC models is essential. 
Higher-order models can represent a more complex building structure, 
yet identifying such models requires more parameters to be estimated by BNN
and increases the amount of training data needed to train the BNN.
In contrast, lower-order models can be learned easily,
but they may not accurately represent the temperature dynamics inside the building.
To decide on the order of the RC model,
we compare the ability of 5 different RC models 
to estimate the indoor temperature for all homes in our dataset.

\subsubsection{Establishing 1R1C models\label{sec:result-rc-length}}
As discussed in Section~\ref{sec:methodology}, 
identifying a 1R1C model is different from other $n$R$n$C models
as the parameters can be estimated directly by solving a non-negative least squares problem.
We use 75 days of data to solve the optimization problem and 
then test it on the remaining 15 days in the same season. 

Our result indicates that in 7,698 homes (out of the 8,884 homes in our dataset),
the obtained parameters are invalid, i.e., the R or C value is zero.
This implies that a meaningful 1R1C model cannot be built for the majority of homes.
We attribute this to the fact that the building envelope 
cannot be accurately modelled by a single thermal resistor.
Interestingly, the 1R1C model achieves a very low RMSE in the remaining 1,186 homes.
The average RMSE is 0.22 degree Fahrenheit if we consider these homes only.
In 120 homes, the obtained 1R1C model could estimate the indoor temperature with 100\% accuracy.
According to the metadata, over 71\% of such homes have the maximum typical occupancy of 2 people or less, 
around 90\% of them have one heating stage and one cooling stage only, 
and around 64\% of them have a floor area of less than 2,000 square foot.
This suggests that the first-order RC model typically performs well 
in small homes with a low occupancy level that have single-stage heating and cooling equipment.
Since the 1R1C model cannot be trained for 86\% of homes using 75 days of training data
and even when it is trained it cannot be transferred to similar homes, 
we turn our attention to other RC models.

\begin{figure}[t!]
\centerline{\includegraphics[width=\columnwidth]{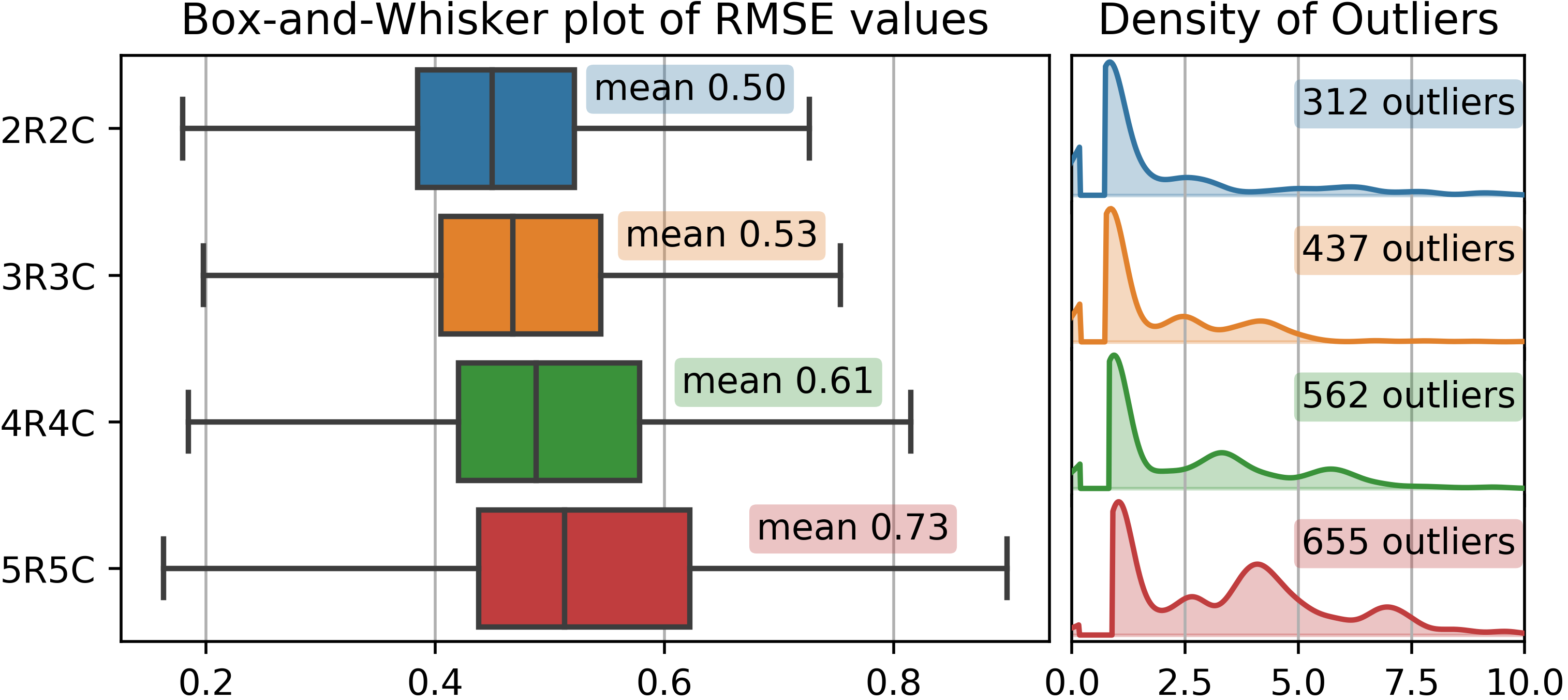}}
\caption{RMSE distribution for different BNN-RC models (left panel) and 
RMSE density for outliers only (right panel). 
An outlier is defined as a home with an RMSE that is greater than 
$1.5\times\text{IQR}$ added to the third quartile or 
is less than $1.5\times\text{IQR}$ subtracted from the first quartile.}
\label{fig:model-order}
\end{figure}

\subsubsection{Selecting the order of the $n$R$n$C model\label{sec:result-rc-order}}
We investigate if higher-order models can give us a better result for most homes. 
We build four different RC models\footnote{We do not consider $n$R$n$C models when $n>5$ 
because it becomes more difficult to train a BNN for estimating parameters of these models 
and that they are not suitable for transfer learning.}, namely 2R2C, 3R3C, 4R4C, and 5R5C,
and estimate their parameters using BNNs.
We use 75 days of data for training the model and test it on the remaining 15 days. 
As shown in Figure~\ref{fig:model-order}
the 2R2C model outperforms higher-order models in terms of RMSE obtained 
in the temperature prediction task.
Considering the 8,884 homes in our dataset, 
2R2C yields a lower average and a lower median RMSE.
Moreover, there are fewer homes with an RMSE that is outside the interquartile range (IQR) multiplied by 1.5.
The RMSE density of these `outliers' is depicted in the right panel of Figure~\ref{fig:model-order}.
Although 4R4C and 5R5C give slightly lower RMSE values than 2R2C for a small number of homes,
their interquartile range is much wider than that of 2R2C 
and their average RMSE is larger too.

Next we examine the RC model that attained the lowest RMSE in each home.
We find that the 2R2C model gives the lowest RMSE in 3,940 homes (44.35\% of homes in our dataset),
while 3R3C, 4R4C, and 5R5C models give the lowest RMSE in
2393 homes (26.94\%), 1525 homes (17.17\%) and 1026 homes (11.55\%), respectively.
This indicates that the best-fit RC model is different for each home, 
but if we are to develop RC models of the same order for all homes in our dataset
we should pick the 2R2C model which is more parsimonious 
and gives a higher prediction accuracy on average than the other models.
Note that the main reason for following this one-size-fit-all approach is that 
it allows for transferring a model built for one home to any other home in the dataset
as we do not need to change the BNN architecture.
Transfer learning is useful when sufficient smart thermostat data is not available for 
the target home to determine the proper order of the RC model and build it from scratch.
In the following we discuss how much data is needed to train an accurate BNN-2R2C model
and how this model can be transferred to other homes or to different seasons.

\subsection{Finding the amount of data needed to train BNN-2R2C models\label{sec:result-rc-length}}
\begin{figure}[t!]
\centerline{\includegraphics[width=\columnwidth]{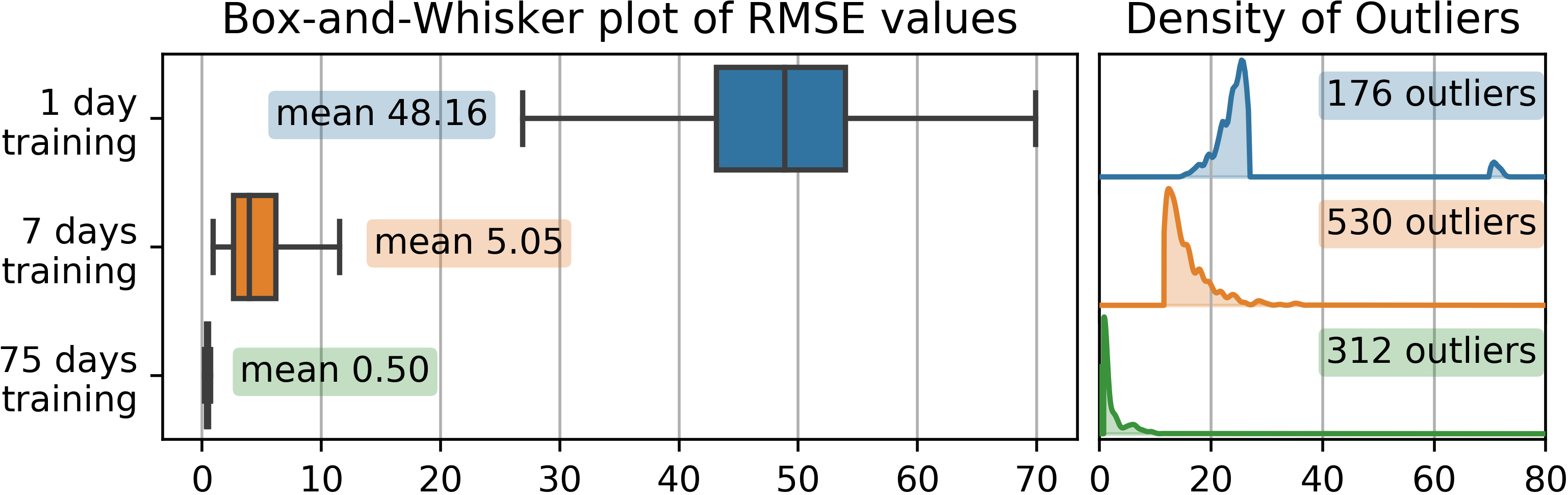}}
\caption{Accuracy of BNN-2R2C trained using different amounts of data.}
\label{fig:training-length}
\end{figure}

We now explore how much data is needed to train a BNN-2R2C that has an acceptable RMSE value.
To this end, we build the BNN-2R2C models using one day, one week, and 75 days of training data 
and test them on the remaining days of the same season.
As can be seen from Figure~\ref{fig:training-length}, 
the models built with 75~days of data performs noticeably better than the other two cases. 
In particular, the models built with 75 days of training data have an average RSME of $0.50$ 
compared to average RSME of $5.05$ and $48.16$ obtained for models 
built with 7~days and 1~day of training data, respectively.
We attribute this to the fact that the BNN needs to learn 11 parameters and 1 bias for the 2R2C model,
hence it needs more than 1~day of training data.
Unfortunately, 75~days (or even 7~days) of training data is not readily available for some homes,
especially the ones that have recently installed a smart thermostat.
This motivates us to take advantage of transfer learning to reduce the amount of training data 
needed for building an accurate 2R2C model.

\subsection{Transfer learning across homes\label{sec:result-transfer}}
As 75 days of training data may not be readily available in a home, 
we try to identify some generic BNN-2R2C models that can be transferred to similar homes. 
We hypothesize that homes with similar floor areas which are constructed around the same time 
should have similar model parameters 
given that they are all located in the same country. 
This is because the heat capacitance of the indoor space shows a strong correlation with the size of the home,
and insulated glazing and multi-layer wall structure are more common among newly built homes.
Based on these observations, we cluster homes according to their floor area and age. 
The advantage of using metadata for clustering is that they are available, 
and once the clusters are formed, they can be used to immediately assign a new home 
to one of these clusters based on some distance measure.

We use k-means clustering and utilize the features mentioned above. 
We run the clustering algorithm on 8,884 homes starting with 2 clusters and 
increasing it to a maximum of 30 clusters.
We find that the sum of squared error (SSE) gradually decreases as 
the number of clusters increases.
The SSE can be defined as:
\[SSE = \sum\limits_{k = 1}^K {\sum\limits_{i = 1}^{{n_k}} {{{({x_{i,k}} - {{\mu}_k})}^2}} }\]
where \(K\) is the total number of cluster, \(n_k\) is the number of members of the \(k\)th cluster, 
\(x_{i,k}\) is the \(i\)th member of the \(k\)th cluster, and \(\mu_k\) is the mean of the \(k\)th cluster.

\begin{figure}[t!]
\centerline{\includegraphics[width=\columnwidth]{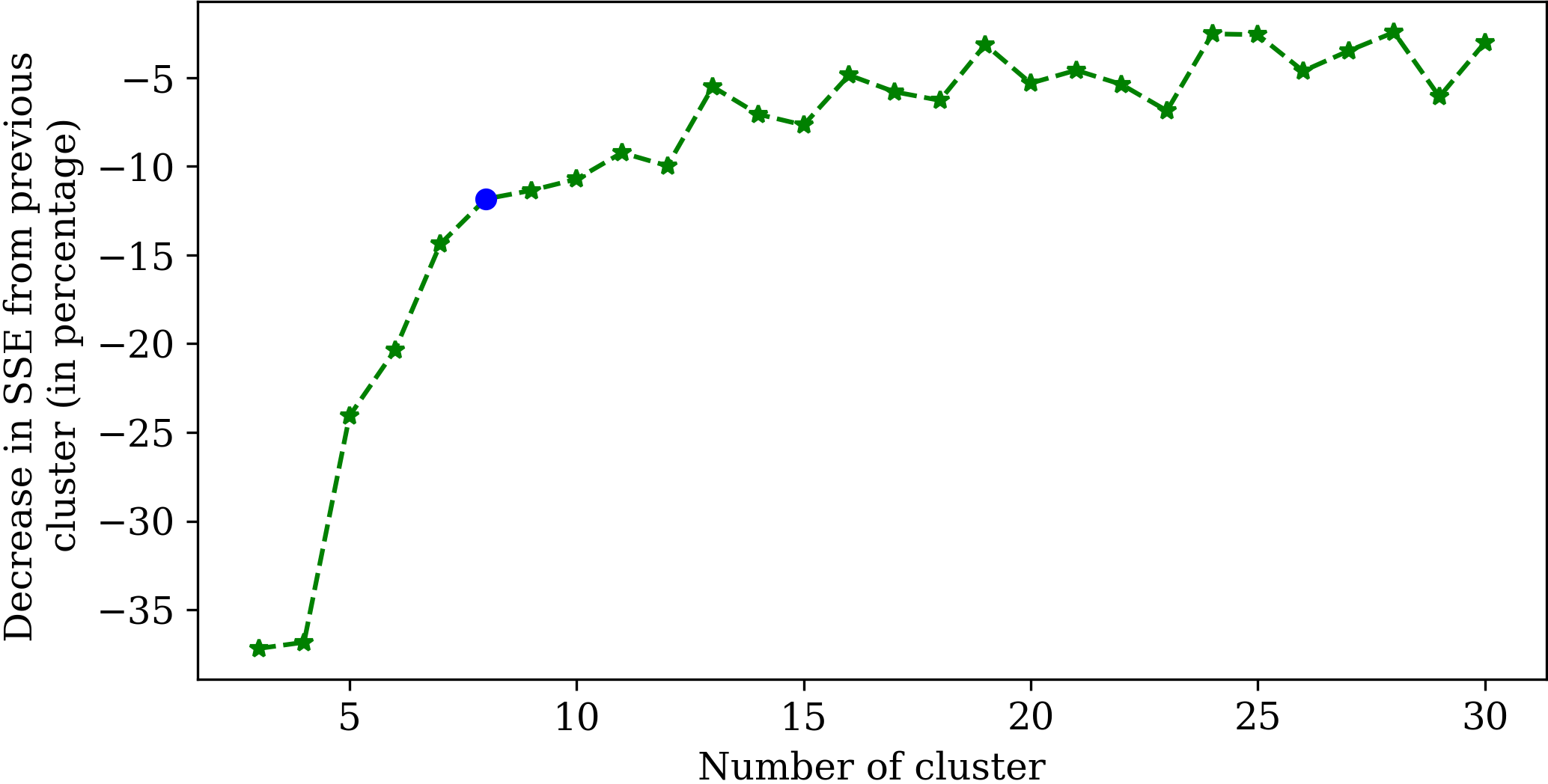}}
\caption{Diminishing return for SSE.}
\label{fig:decrease}
\end{figure}

We use the elbow method to determine the number of clusters.
We plot the diminishing return for the SSE values which can be written as:
\[d = \frac {SSE_{\kappa+1}-SSE_{\kappa}}{SSE_{\kappa}}\times 100\] 
As it can be seen from Figure~\ref{fig:decrease}, the comparative decrease becomes flat after 8 clusters. 
Therefore, we set the number of clusters to 8 and assign every home to its corresponding cluster.
Since k-means calculates the cluster centre based on the arithmetic mean, 
the centre does not necessarily represent a real home. 
We identify the closest real homes to the k-means cluster centres and 
treat them as representative homes of their clusters for transfer learning.
We transfer the BNN-2R2C model of the representative home of each cluster
which is trained using sufficient training data (75 days) 
to other members of that cluster which we refer to as target homes.

\begin{figure}[t!]
\centerline{\includegraphics[width=\columnwidth]{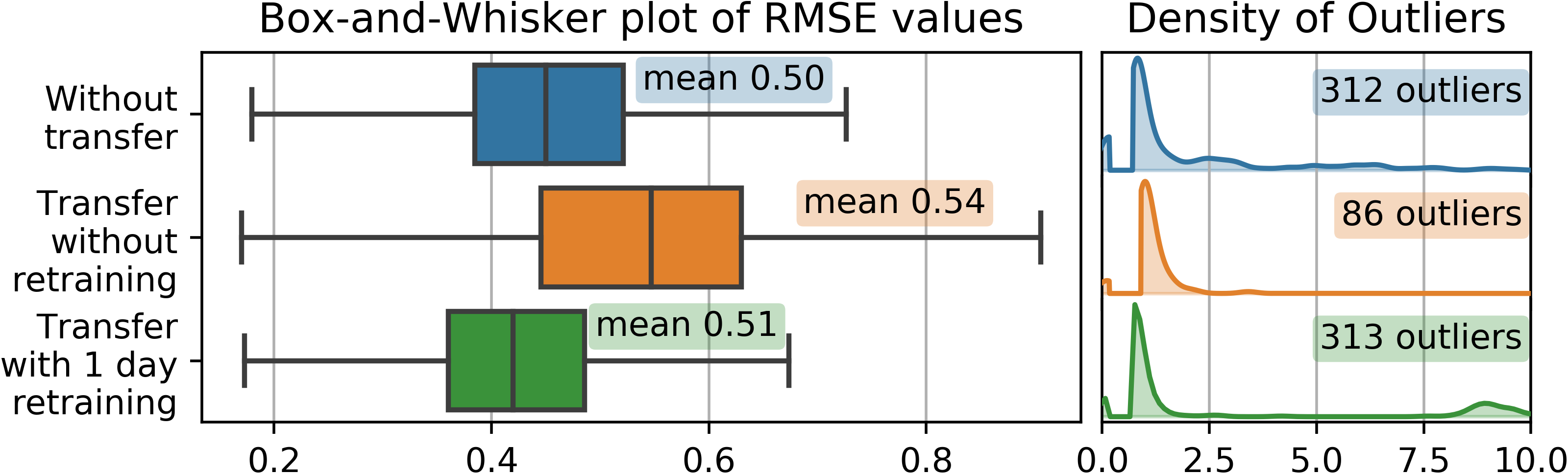}}
\caption{Comparison of RMSE when the model is trained for each home using 75 days of training data
and when the pre-trained model of the representative home of the respective cluster 
is transferred to each home with and without retraining.}
\label{fig:transfer}
\end{figure}

It can be seen from Figure~\ref{fig:transfer} that the performance of 
the transferred BNN-2R2C model is comparable with
that of the BNN-2R2C model that could have been built from scratch 
if 75 days of smart thermostat data was available from the target home.
The average RMSE is 0.54 degree Fahrenheit when we transfer directly, 
i.e., we do not retrain the transferred source model using time series data from the target home. 
We get a slightly better average RMSE of 0.51 
if we use 1~day data from target homes to retrain the transferred model.
That said, even without adaptation, we get almost the same average RMSE as we got if we had 75~days of training data.
This suggests that we can simply transfer the thermal model of the representative home of that cluster
to a home that recently installed a smart thermostat
and use this model for optimal control from the first day that the smart thermostat is installed.

\subsection{Transfer learning across seasons\label{sec:result-transfer-season}}
Our experiments show that the RC model trained in one season 
does not typically achieve the same level of prediction accuracy when used in another season.
This is in line with what has been reported in~\cite{Pathak2019}
and can be attributed to latent variables that we did not capture in our model 
or changes in the effective RC parameters from one season to another.
Nevertheless, it is imperative to update or retrain the thermal model over time.
To this end, we transfer the individual and cluster representative models trained 
for winter to summer to evaluate its performance.

\begin{figure}[t!]
\centerline{\includegraphics[width=\columnwidth]{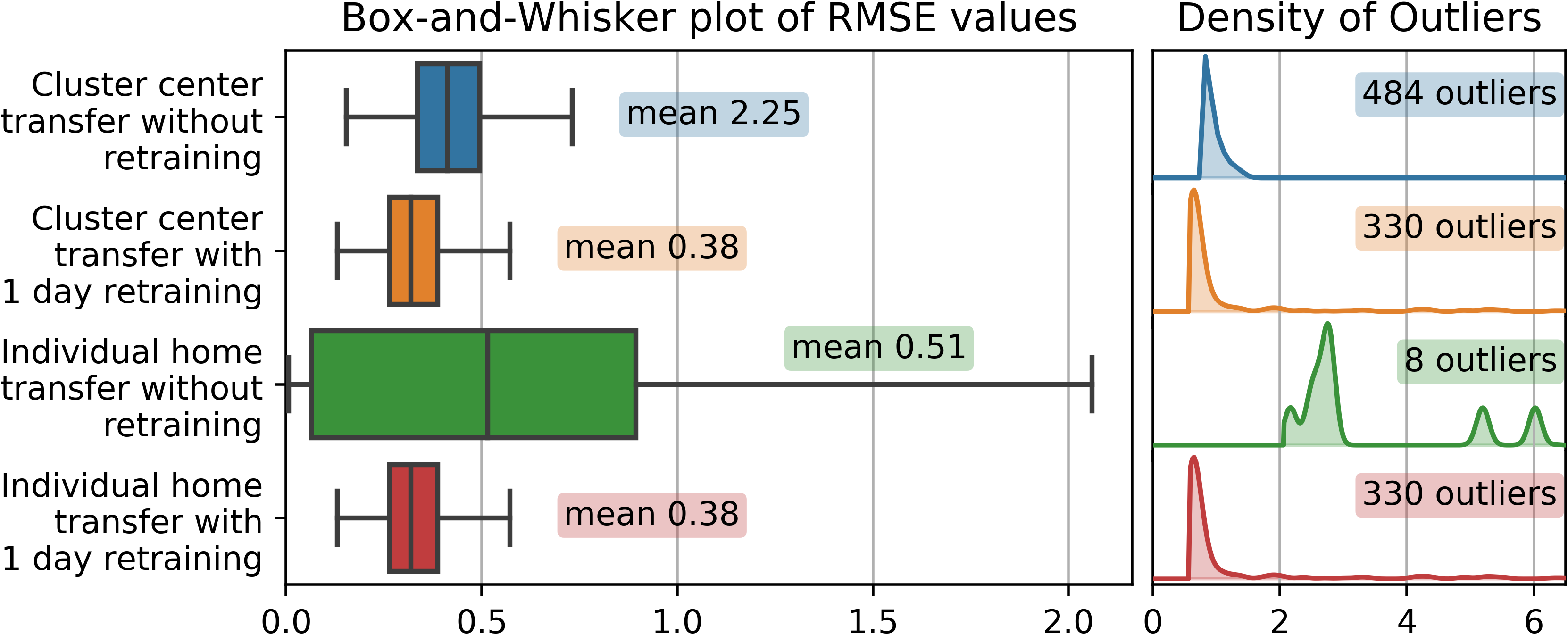}}
\caption{Comparison of RMSE distributions when transferring models to the summer season.}
\label{fig:summer}
\end{figure}

The homes in our dataset have different lengths of time series data and 
not all of them include data from both seasons. 
Thus, in this experiment we only consider homes that had data for both summer and winter. 
We consider two scenarios. 
The first scenario is where the winter model of the representative home of the respective cluster 
is transferred and used (with and without retraining) as the summer model in the target home.
The second scenario is where the winter model developed for the target home using 75 days of data
is transferred to summer with and without retraining. 
In both scenarios, time series data from 1~day in summer is utilized when the model is retrained.
Data from the remaining days in the summer season is used to test the models.
Figure~\ref{fig:summer} shows that in both scenarios, 
we obtain mean RMSE of 2.25 and 0.51 without retraining,
which is markedly high in the case that models of representative homes are transferred.
The mean RMSE and its variance decrease considerably using 1 day of retraining data from summer. 
This shows that retaining is necessary when transferring across seasons, 
especially if the transferred model does not belong to the same home.
Another observation is that if adaptation is performed using 1~day of data from the target home, 
transferring the pre-trained winter model of the representative home
is as effective as transferring the accurate winter model that is trained for the same home.

\subsection{Comparing RC models and black-box models}
Bayesian neural network is not the only model that can estimate the parameters for the RC model. 
A standard neural network (which uses point estimates unlike BNN) is also capable of estimating 
the parameters in Equation~(\ref{eq:rc-solution}).
To justify the need for a BNN to learn the model parameters,
we estimate parameters of a 2R2C model using a neural network; we refer to this model as NN-2R2C.
Our experiments suggest that this model performs poorly compared to the BNN-2R2C model.
In particular, assuming that both models are trained 75 days of smart thermostat data,
the average RMSE is 25.87 for NN-2R2C and 0.50 for BNN-2R2C. 
The uncertainty introduced by the BNN-2R2C model addresses the overfitting problem. 
Moreover, the Bayesian approach offers other advantages when it comes to building an RC model, 
e.g., it enables us to incorporate the prior knowledge regarding the model parameters
when estimating model parameters in the target domain. 

\begin{figure}[t!]
\centerline{\includegraphics[width=\columnwidth]{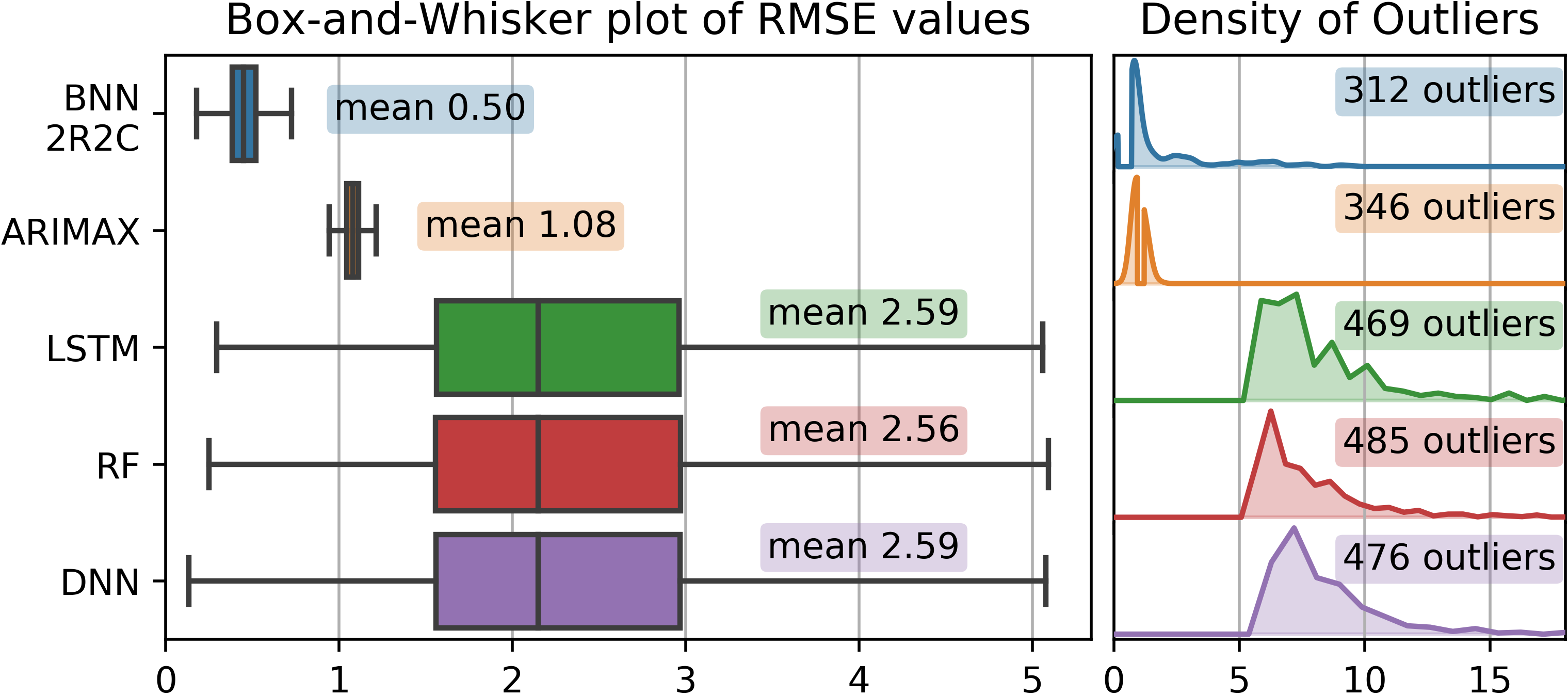}}
\caption{Comparison of RMSE distributions of BNN-2R2C and baseline black-box models.}
\label{fig:black-gray}
\end{figure}

We also benchmark the BNN-2R2C model with other black-box models 
introduced in Section~\ref{sec:methodology}, including ARIMAX, LSTM, RF, and DNN. 
All these models are trained using 75 days of data and tested using 15 days of data from the same season. 
Figure~\ref{fig:black-gray} shows the performance of all models 
when used to predict the indoor temperature.
As it can be seen, the BNN-2R2C outperforms all black-box models and has an average RMSE of 0.50.
The ARIMAX model is the second best model with an average RMSE of 1.08 and a narrow spread of RMSE values. 
The BNN-2R2C is superior to the ARIMAX model (in terms of RMSE) in 94.4\% of homes.
LSTM, RF, and DNN give an average RMSE of 2.59, 2.56, and 2.59, respectively. 
This shows the efficacy of the proposed method for building accurate thermal models.

\section{Conclusion}
This paper studies the problem of identifying grey-box thermal models (RC-network models) 
with Bayesian neural networks leveraging time series data generated by smart thermostats 
and metadata about the homes.
These models have superior performance in estimating the indoor temperature;
thus, they are suitable for model-based control of heating and cooling equipment. 
We argued that since building accurate grey-box models 
requires at least several days of training data,
a library of pre-trained thermal models from representative homes 
can be built and one model from this library can be chosen and transferred to the target home to achieve high accuracy.
The representative homes are selected via clustering of the metadata that is available in our dataset.

Using real data collected by ecobee smart thermostats installed in over 8,000 homes in Canada,
we investigated which order of the developed RC model can better
describe the heat flux in the many homes in our dataset.
We found that on average the 2R2C model can perform better than other $n$R$n$C models.
We showed that it is crucial to use BNN to estimate the parameters of the RC model, and 
compared the performance of the BNN-2R2C model with various black-box thermal models proposed in the literature. 
Furthermore, we explored the idea of transferring the BNN-$n$R$n$C model 
across seasons in the same home and across homes that have similar characteristics.
Transfer learning can greatly reduce the need for training data and 
would ensure achieving higher accuracy targets in estimating the indoor temperature.

One limitation of this work is that we cannot uniquely identify the $R$ and $C$ parameters of an $n$R$n$C model
given the compound $RC$ parameters estimated by the Bayesian neural network.
We have to at least know the true value of $Q_{heat}$ or $Q_{cool}$ to solve for $R$ and $C$,
but this information is not included in the ecobee dataset.
Should we know the amount of heat flux from the HVAC system, 
these parameters can be uniquely identified
and possibly utilized to conduct virtual energy audits, detect faults, and offer energy saving recommendations.

In future work we plan to incorporate the grey-box models built for the homes in our dataset
to implement various control algorithms.
Using a co-simulation platform, we will simulate the resulting control policy 
(i.e., adjust the temperature setpoints over time)
to calculate the HVAC energy use and study thermal comfort.
This enables us to compare the BNN-$n$R$n$C model with other competing thermal models 
in terms of potential energy savings and impact on occupant comfort.


\bibliographystyle{ACM-Reference-Format}
\bibliography{common}
\end{document}